\documentstyle[epsfig]{mn}
\begin{document}
\input BoxedEPS.tex
\newcommand{\aip}{{\small ${\cal AIPS}$}}
\newcommand{\gtsim}{\mbox{{\raisebox{-0.4ex}{$\stackrel{>}{{\scriptstyle\sim}}
$}}}}
\newcommand{\ltsim}{\mbox{{\raisebox{-0.4ex}{$\stackrel{<}{{\scriptstyle\sim}}
$}}}}
\newcommand{\s}{$\stackrel{\rm s}{.}$}
\newcommand{\h}{$^{\rm h}$}
\newcommand{\m}{$^{\rm m}$}
\newcommand{\pp}{$\stackrel{\prime\prime}{.}$}
\newcommand{\de}{$^{\circ}$}
\newcommand{\p}{$^{\prime}$}
\newcommand{\arc}{$^{\prime\prime}$}
\newcommand{\marc}{^{\prime\prime}}
\newcommand{\rs}{{\em $r_s$}}
\newcommand{\DPM}{{\em DPM}}
\newcommand{\alf}{{\displaystyle\biggl({\nu_{\rm h} \over \nu_{\rm l}}\biggr)^{\alpha}} }

\newcommand{\figstart}[1]
    { \begin{figure}[htb]
      \begin{picture}(0,#1) }
\newcommand{\figend}[4]
    { \end{picture}
      \special{#1}
      \caption[#2]{#3}
      \label{#4}
      \end{figure} }
\newcommand{\fig}[5]
    { \figstart{#1}
      \figend{#2}{#3}{#4}{#5} }
\newcommand{\bHS}{\beta_{\mbox{\scriptsize HS}}}
\newcommand{\bBF}{\beta_{\mbox{\scriptsize BF}}}
\newcommand{\nT}{\nu_{\mbox{\scriptsize T}}}
\newcommand{\et}{E_{\mbox{\scriptsize T}}}
\newcommand{\nTn}{\nu_{\mbox{\scriptsize Tn}}}
\newcommand{\nTf}{\nu_{\mbox{\scriptsize Tf}}}
\newcommand{\tn}{\tau_{x\mbox{\scriptsize n}}}
\newcommand{\tf}{\tau_{x\mbox{\scriptsize f}}}
\newcommand{\xn}{x_{\mbox{\scriptsize n}}}
\newcommand{\xf}{x_{\mbox{\scriptsize f}}}
\newcommand{\yn}{y_{\mbox{\scriptsize n}}}
\newcommand{\yf}{y_{\mbox{\scriptsize f}}}
\newcommand{\lln}{l_{\mbox{\scriptsize n}}}
\newcommand{\llf}{l_{\mbox{\scriptsize f}}}
\newcommand{\Dn}{f(\Delta_{\mbox{\scriptsize n}})}
\newcommand{\Df}{f(\Delta_{\mbox{\scriptsize f}})}
\newcommand{\B}{\mbox{$B$}}
\newcommand{\Bo}{\mbox{$B$}_{0}}

\SetRokickiEPSFSpecial
\HideDisplacementBoxes

\title[]{Cirrus models for local and high z SCUBA galaxies}
\author[]{Andreas Efstathiou$^{1}$ and Michael Rowan-Robinson$^2$\\
$^1$ Department of Computer Science  \& Engineering, Cyprus College, 6
Diogenes Str, 1516 Nicosia, Cyprus. 
\\
$^2$ Astrophysics Group, Blackett Laboratory, Imperial College of Science Technology and Medicine,
Prince Consort Road, London SW7 2BZ
}
\maketitle
\begin{abstract}

We present a model for the UV to submillimeter emission from stars
embedded in the general interstellar dust in galaxies (the 'infrared
cirrus' component).  Such emission is characterized by relatively low
optical depths of dust and by cool ($<$ 30 K) dust temperatures.  The
model incorporates the stellar population synthesis model of Bruzual
\& Charlot and the dust model of Siebenmorgen \& Kr\"{u}gel which
includes the effects of small grains/PAHs.  We apply the model to fit
the optical to submillimeter spectral energy distributions (SEDs) of
nearby galaxies which are dominated by cirrus emission and find that
our simple model is quite adequate to explain the observed SEDs.

We also, more controversially, apply this cirrus model to the SEDs of
high redshift sources detected in blank field submillimeter surveys
with SCUBA.  Surprisingly, an excellent fit is found for many of these
sources, with typical values for the optical depth $A_V$ and the
surface brightness of the stellar radiation field $\psi$ being only a
 factor 2-3 higher than for nearby galaxies.  This increase is not
unreasonable given the expected evolution of dust optical depth in
currently favoured star-formation history models.

We conclude that the tendency to interpret the high-z SCUBA galaxies
as very highly obscured starbursts may be premature and that these
galaxies may be more closely linked to optically selected high
redshift galaxies than previously assumed.

\end{abstract}



\section{Introduction}

The need for several different components to understand the infrared
and submillimetre spectral energy distributions (SEDs) of galaxies has
been recognized for some time.  Rowan-Robinson and Crawford (1989)
modelled IRAS galaxy colour-colour diagrams in terms of three
components, starburst, cirrus and AGN components.  The cirrus
component corresponded to the absorption of the general starlight of a
galaxy by interstellar dust and reemission in the infrared, as
identified in our Galaxy by Low et al (1984). Detailed models for the
cirrus component were given by Rowan-Robinson (1992), but these did
not include a treatment of the line emission from the PAH component.
A model of the infrared emission of the solar neighbourhood that
included the effect of small grains and PAHs was presented by
Siebnmorgen \& Kr\"{u}gel (1992).  Efstathiou, Rowan-Robinson \&
Siebenmorgen (2000; hereafter ERS2000) have given the most detailed
model to date of the evolution of the infrared emission from a
starburst.  In this paper we describe a tool which allows the cirrus
emission to be modelled from 0.1-1000 $\mu$m and apply it both to
local galaxies which have been mapped with SCUBA and to high redshift
galaxies found in deep SCUBA surveys at 850 $\mu$m.  Hitherto there
has been a tendency to interpret these SCUBA galaxies as high redshift
versions of Arp220, but Rowan-Robinson (2001) has pointed out that
submillimetre counts and background radiation are much more naturally
understood in terms of cirrus-like emission.

Silva et al (1998) and Granato et al (2000) have given models for
infrared SEDs which include both starburst and cirrus
components. These models take into account the distribution of dust
and stars in the galaxy and incorporate a population synthesis
model. The effect of changes in metallicity during the lifetime of a
galaxy is also taken into account.  The cirrus models we present here
are simpler than those of Silva et al (1998) and Granato et al (2000)
but appear to be just as good a representation of the observed SEDs of
nearby normal galaxies. Starburst models that take into account the
distribution of dust and stars were also presented by Kr\"{u}gel 
\& Siebenmorgen  (1994) and Siebenmorgen, Kr\"{u}gel \& Laureijs (2001).

The layout of this paper is as follows: section 2 describes our cirrus
model in detail, section 3 applies this model to nearby galaxies which
have been mapped with SCUBA, section 4 discusses the application to
high redshift SCUBA galaxies, and section 5 gives our discussion and
conclusions.

\section{The cirrus model}

\begin{figure*}
\epsfig{file=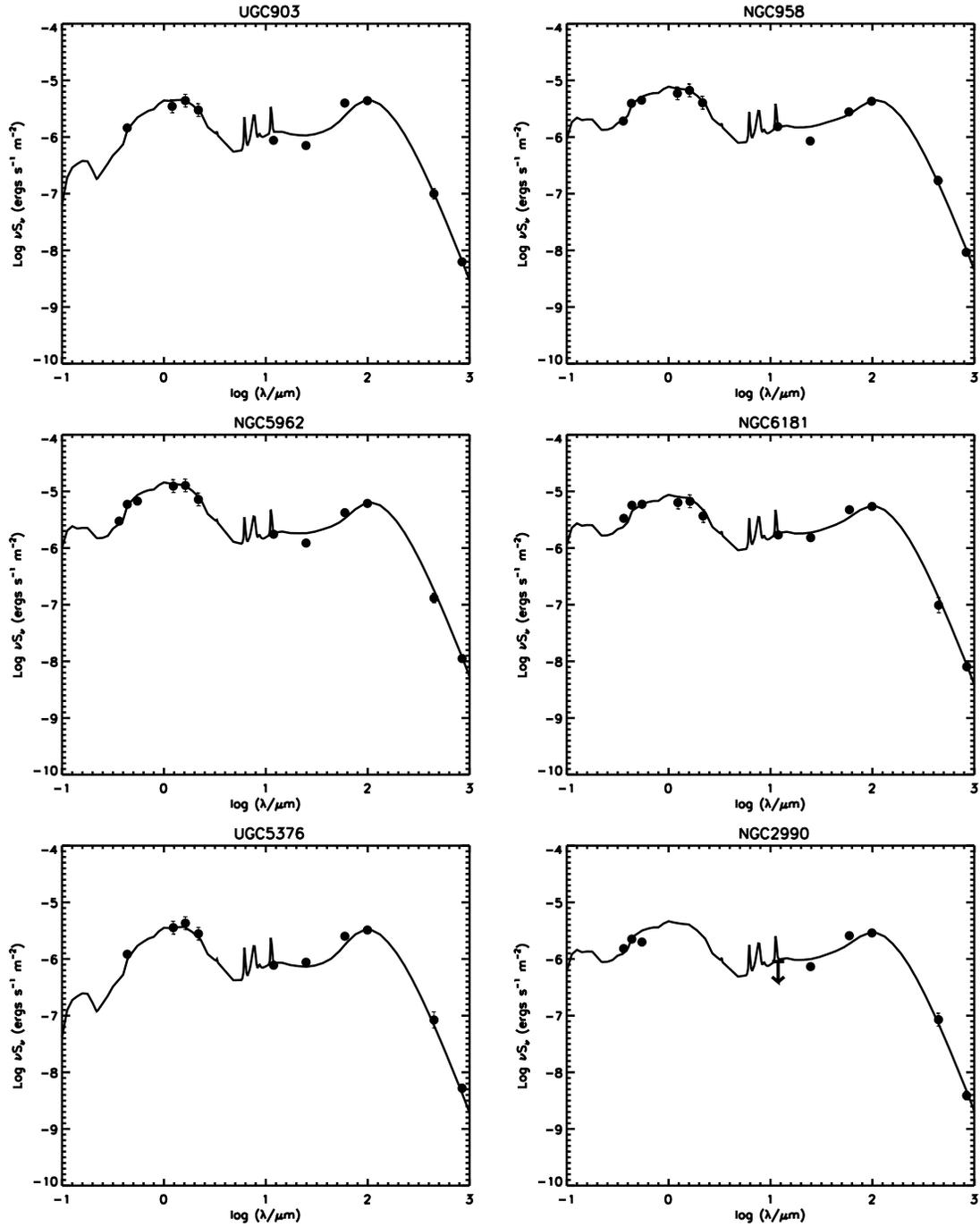, angle=0, width=15cm}
\caption{
Spectral energy distributions of nearby cirrus galaxies. Data from
IRAS, Dunne \& Eales (2001) and the NASA/IPAC Extragalactic Database (NED).
Model parameters given in Table 1.
}\label{nearby}
\end{figure*}

\begin{table*}
\begin{tabular}{llrll}
Name             & $A_V$      & $\tau$/Gyrs  &$\psi$ &  $\chi^2$/ndf     \\
                 &            &              &       &                   \\
UGC903           &  0.9       &   6          &  4    &  1.05             \\
NGC958           &  0.4       &   9          &  2    &  0.56             \\
NGC5962          &  0.4       &   5          &  3    &  0.56             \\
NGC6181          &  0.4       &   11         &  3    &  0.97             \\
UGC5376          &  0.9       &   5          &  5    &  0.35             \\
NGC2990          &  0.4       &   11         &  3    &  1.33             \\
NGC1667          &  0.5       &   11         &  8    &  1.42             \\
\end{tabular}
\caption{\label{tab:sample}
Fitted parameters for the nearby cirrus galaxies: dust optical depth, $A_V$;
exponential time-scale for star formation, $\tau$; intensity of interstellar
radiation field relative to solar neighbourhood, $\psi$; goodness of fit,
$\chi^2$.  All models assume a galaxy age $t_*$ of 12.5Gyrs, $f=1$ and
$t_m=3$Myrs. 
}\end{table*}

The ingredients of a cirrus model are (1) the input stellar radiation
field, which has to be characterized both with regard to its spectrum
and its intensity, (2) an interstellar dust model, with assumptions
about its distribution and opacity, (3) a radiative transfer treatment
of the interaction between the two to generate the output SED.  Our
goal is to provide a versatile tool which can be used for example in
simulations of the galaxy formation history.

For specifying our input stellar radiation field we have used the
Galaxy Isochrone Synthesis Spectral Evolution Library (GISSEL, Bruzual
\& Charlot 1993) that gives the ultraviolet to  near infrared spectrum of
radiation from a mass of stars as a function of time.

Stars form in dusty molecular clouds and spend a considerable amount
of time inside them. During this phase, the stellar radiation is
absorbed by molecular cloud dust and reprocessed to the infrared. We
use the code of ERS2000 to compute the emission of the stellar
population during this phase. The molecular cloud is assumed to
disperse $7.2 \times 10^7$ years after the instantaneous formation of
stars inside it. However, well before $7.2 \times 10^7$ years
non-spherical evolution of the molecular cloud may allow a fraction
$f$ of the starlight to escape without any dust absorption. We assume
that this occurs at time $t_m$ after star formation. By further
assuming that the radiation field in a galaxy is due to a large number
of randomly oriented molecular clouds, their average emission (for
stars in the age range $t_m$ to $7.2 \times 10^7$ years) is
approximately

$$  (1-f) S_\nu^s  + f S_\nu^* $$
where $S_\nu^s$ is the emission from a spherical GMC (as computed by
ERS2000) and $S_\nu^*$ is the emission from the stellar population in the
absence of any molecular cloud dust (Bruzual \& Charlot). The emission
of stars younger than $t_m$ or older than $7.2 \times 10^7$ years is
assumed to be $S_\nu^s$ or $S_\nu^*$ respectively. In fact for most
of this paper we have simplified the model by assuming $f=1$. The effect
of allowing $f < 1$ is discussed in section 5.

 Other parameters
which have to be specified are the star formation history function
$\dot{\phi}_*(t)$, the epoch at which this history begins, $t_i$, the
epoch of observation, $t$, and the usual parameters for the stellar
IMF.  Since the quantity which determines the dust temperature is the
intensity of the stellar radiation field, we have chosen to
characterise this in terms of the ratio of the bolometric intensity of
the radiation field to that in the stellar radiation field in the
solar neighbourhood (Mathis et al 1983), $\psi$.  The star-formation
rate, $\dot{\phi}_*(t)$, is assumed to have an exponential form, with
exponential time scale, $\tau$, and a Salpeter IMF from 0.1-125
$M_{\odot}$ is assumed.  The age of the galaxy, $t_*\equiv t-t_i$,
 can also be specified.

We use the dust grain model of Siebenmorgen and Kr\"{u}gel (1992),
which incorporates a detailed treatment of small grains and PAHs, and
then characterize the opacity of the interstellar dust by $A_V$.
$A_V$ determines how much of the uv to near-ir light is absorbed by
dust and reemitted in the infrared and submillimetre bands, and
therefore controls the ratio of luminosity in the far infrared to the
luminosity in the uv to near-ir.  If $A_V <<$ 1, we are in the
optically thin regime and the dust temperature will then be determined
by $\psi$ and the shape of the input spectrum (determined by the star
formation history).  For $A_V \ge 1$ but not $>> 1$, the dust will
modify the emergent optical and uv spectrum significantly, but will
still not result in any significant change to the dust temperature
along the line of sight to the central plane of the galaxy.  We make a
further simplifying assumption, that for the purposes of modelling the
galaxy's emergent SED, we can characterize the galaxy by a single
average value of $\psi$.  Strictly speaking we should model the
density distribution of both dust and stars and carry out a full
radiative transfer calculation to compute the emergent spectrum.  The
emergent spectrum would essentially be a superposition of spectra
corresponding to different values of $\psi$ at different locations
through the galaxy.  In practice we find that a single value of $\psi$
characterizes the emergent spectra very adequately.

To summarize, the parameters of the model that have to be set are
$\psi, \tau, t_*, A_V, f$ and $t_m$. The parameters $\tau, t_*, f$ and
$t_m$ determine the shape of the optical SED, with $A_V$ determining
the reddening to be applied to this and also, in a self-consistent
manner, the bolometric luminosity of the infrared emission. $\psi$
determines the temperature of the different grain species in the model
and hence the shape of the far infrared SED. There is also a slight
dependence of grain temperature on the shape of the optical SED. For
$t_*=0.25$ Gyrs and $\tau=6$ Gyrs the temperature of the large
classical grains
is in the range 17-24K for $\psi =1$ and 28-41K for $\psi =21$. Note
that the far-infrared emission is dominated by the emission of the
largest and coolest grains.  For $t_*=12.5$ Gyrs and $\tau=6$ Gyrs the
corresponding temperature ranges are 16-21K for $\psi =1$ and 26-36K
for $\psi =21$.

Our code and a selection of models is available at  
http://astro.ic.ac.uk/$\sim$ane/cirrus-models.html

\section{Application to nearby galaxies}

\begin{figure*}
\epsfig{file=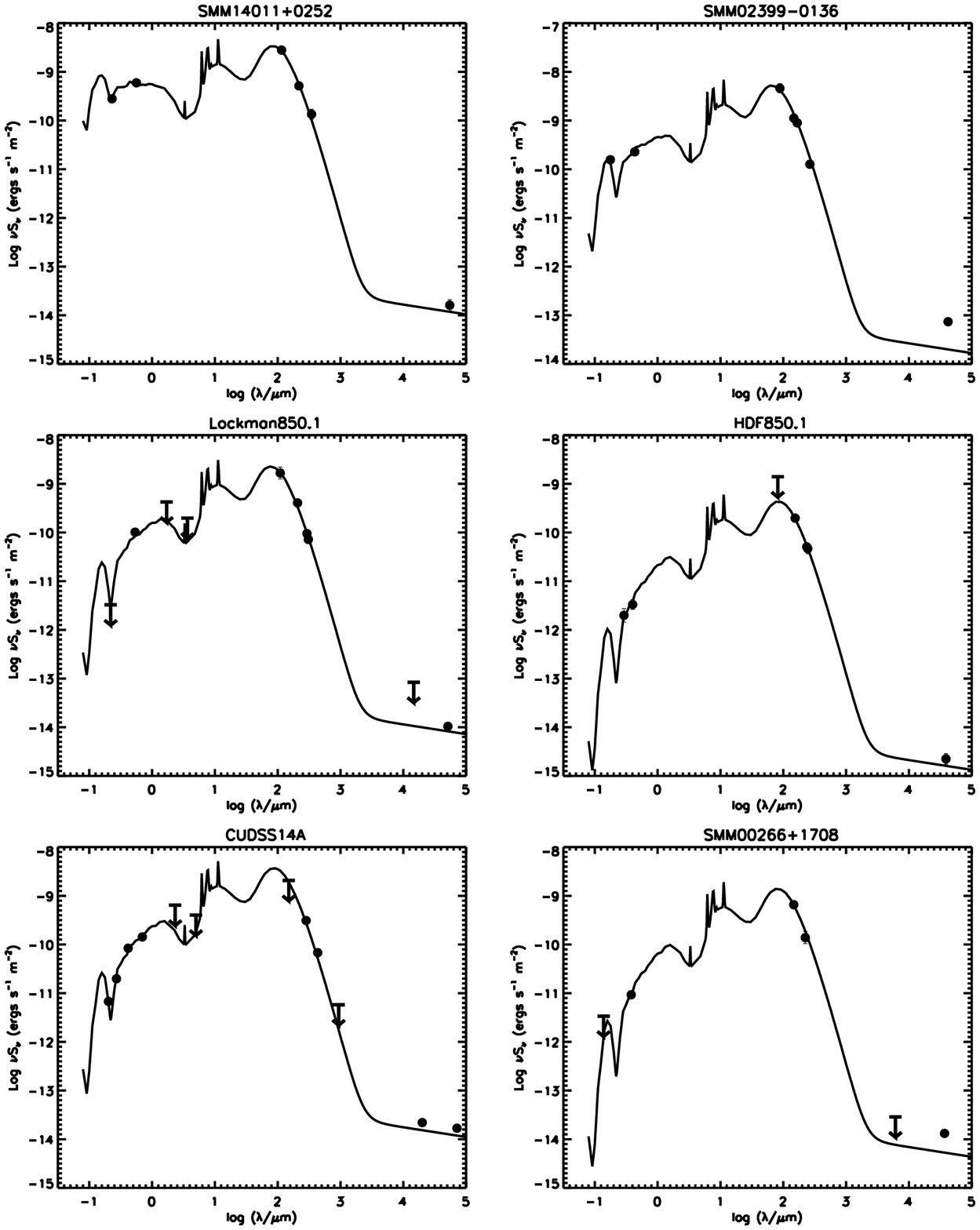,angle=0,width=15cm}
\caption{
Spectral energy distributions of high redshift cirrus galaxies
confirmed by millimetre interferometry. Data from Lutz et al (2001), Dunlop
et al (2002), Ivison et al (1998, 2000, 2001), Frayer et al (2000)
 and Gear et al (2000).  Model parameters given in Table
2. For HDF850.1 only the $z=4.5$ fit is shown.
}\label{nearby}
\end{figure*}

\begin{figure*}
\epsfig{file=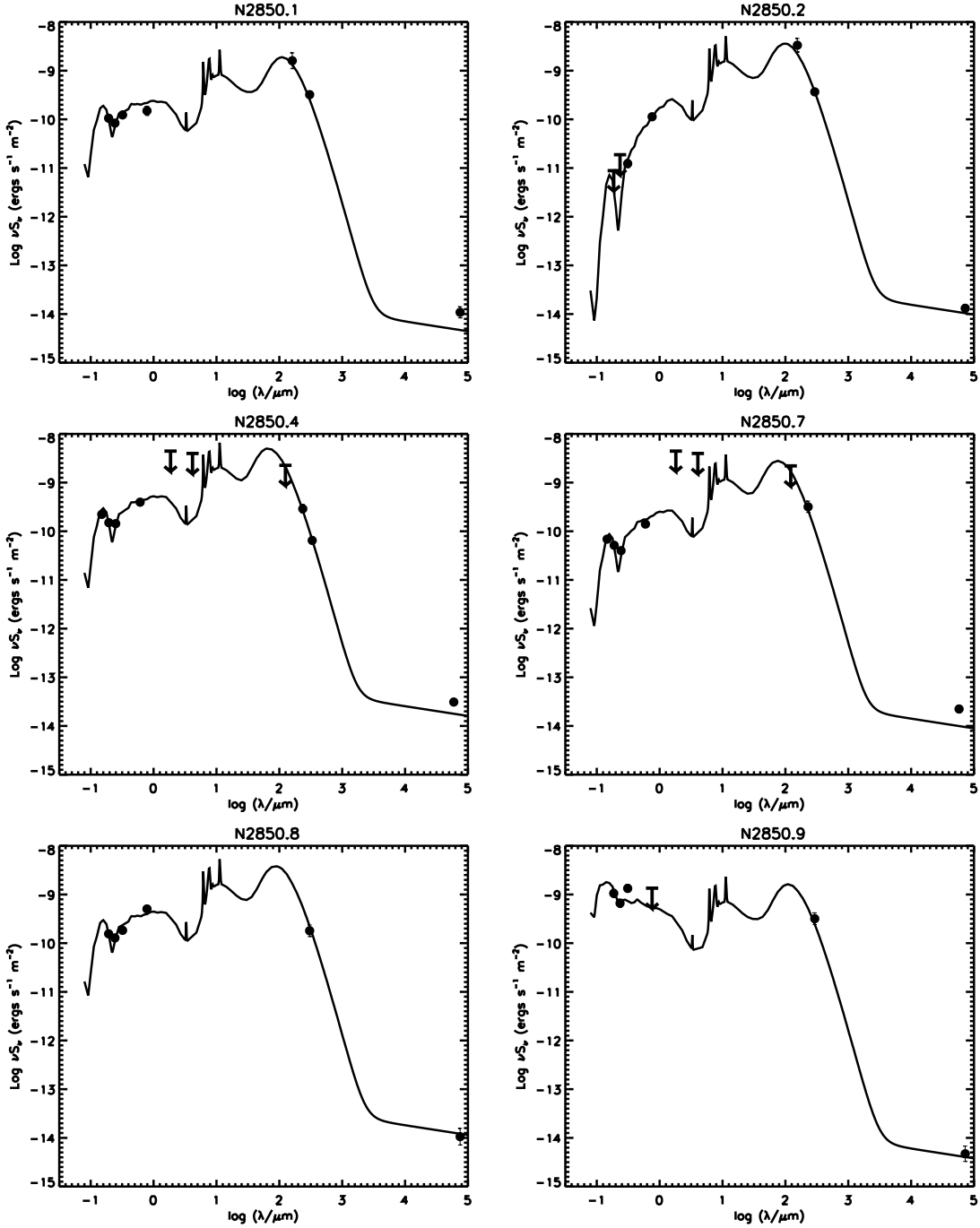,angle=0,width=15cm}
\caption{
Spectral energy distributions of high redshift cirrus galaxies from the 8 mJy
SCUBA survey in ELAIS-N2. Data from
Fox et al(2001), Ivison et al (2002).  Model
parameters given in Table 2. 
}\label{nearby}
\end{figure*}

\begin{figure*}
\epsfig{file=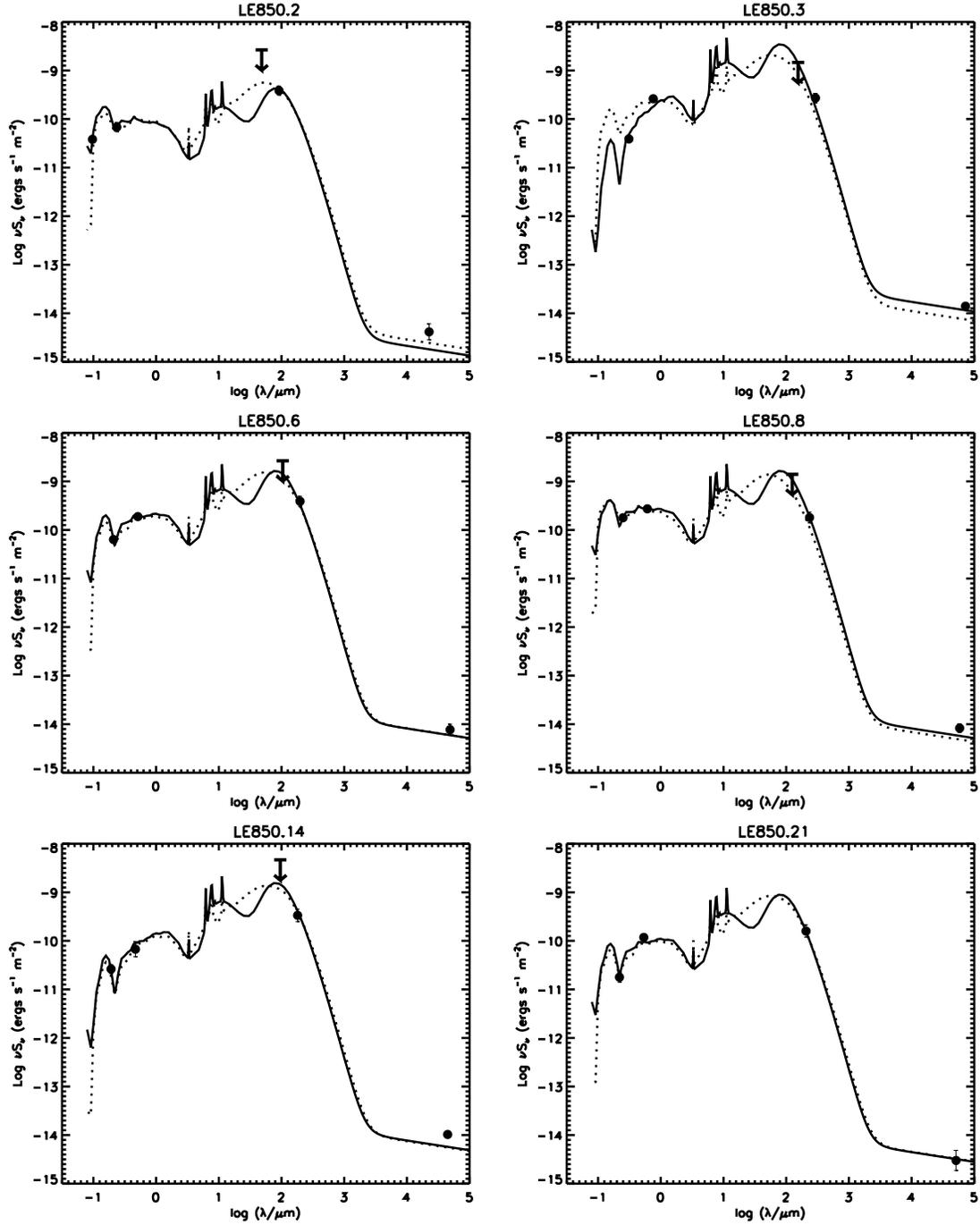,angle=0,width=15cm}
\caption{
Spectral energy distributions of high redshift cirrus galaxies from the 8 mJy
SCUBA survey in Lockman. Data from
Fox et al(2001), Ivison et al (2002).  Model
parameters given in Table 2. Fits with a cirrus model that includes
 a dust enshrouded phase, discussed
in section 5, are shown with the dotted lines.
}\label{nearby}
\end{figure*}

\begin{table*}
\begin{tabular}{lrrllllr}
Name           &$\chi^2/ndf$ & $\psi$ & $A_V$ &  $z$ &  lensing & $log_{10}$  ${{L_{bol}}\over {L_\odot}}$  &  SFR      \\
               &  &        &    &      &   magn.            &   & $M_\odot$/yr            \\
&&&&&&& \\
Objects with mm interferometry &&&&&&& \\
&&&&&&& \\

 SMM14011+0252&   0.824& 5& 0.9&$2.9^{+0.5}_{-0.5}$&3&   12.76& 731\\
 SMM02399-0136&  14.676&21& 1.8&$4.1^{+0.1}_{-0.1}$&2.5& 13.31&2577\\
 Lockman850.1 &   3.329& 8& 2.3&$3.1^{+0.1}_{-0.1}$&-&   13.05&1441\\
    HDF850.1  &   0.638& 5& 3.0&$4.5^{+0.3}_{-0.2}$&3&   12.23& 219\\
              &   1.760&21& 2.1&$7.1^{+0.2}_{-0.1}$&3&   12.53& 436\\
    CUDSS14A  &   1.410& 4& 2.5&$2.0^{+0.1}_{-0.1}$&-&   12.81& 819\\
 SMM00266+1708&   7.840&(6)& 3.1&$4.8^{+0.1}_{-0.2}$&2.4&12.91&1033\\

 &&&&&&& \\
ELAIS-N2 objects  &&&&&&& \\
 &&&&&&& \\

     N2850.1&   3.923& 1& 1.3&$1.8^{+0.1}_{-0.1}$&-&   12.45& 360\\
     N2850.2&   3.041& 2& 3.1&$1.9^{+0.1}_{-0.1}$&-&   12.75& 705\\
     N2850.4&   2.007&20& 1.5&$2.6^{+0.1}_{-0.1}$&-&   13.22&2121\\
     N2850.7&   6.122& 9& 1.8&$2.7^{+0.1}_{-0.1}$&-&   13.01&1295\\
     N2850.8&   8.378& 4& 1.4&$1.8^{+0.1}_{-0.1}$&-&   12.74& 700\\
     N2850.9&   7.789& 1& 0.4&$1.9^{+0.1}_{-0.1}$&-&   12.63& 542\\

 &&&&&&& \\
LE objects &&&&&&& \\
 &&&&&&& \\

     LE850.2&   0.936& (6)& 0.7&$8.3^{+0.4}_{-4.3}$&-&   13.46&3625\\
     LE850.3&  23.859& (6)& 2.3&$1.9^{+1.1}_{-0.6}$&-&   12.72& 673\\
     LE850.6&   0.777& (6)& 1.2&$3.3^{+0.2}_{-0.1}$&-&   13.02&1343\\
     LE850.8&   0.275& (6)& 0.9&$2.6^{+0.4}_{-0.2}$&-&   12.81& 815\\
    LE850.14&   2.507& (6)& 1.8&$3.7^{+0.3}_{-0.1}$&-&   13.09&1566\\
    LE850.21&   3.593& (6)& 1.3&$3.1^{+0.1}_{-0.1}$&-&   12.69& 620\\
     LE850.7&  71.613& (6)& 1.3&$3.0^{+0.1}_{-0.1}$&-&   13.27&2366\\

 &&&&&&& \\
LE objects  &&&&&&& \\
cirrus model with starburst phase ($f=0.5$, $t_m=10^7$yrs) &&&&&&& \\
 &&&&&&& \\

     LE850.2&   0.207& (6)& 0.3&$7.9^{+0.2}_{-0.4}$&-&   13.52&4222\\
     LE850.3&  10.838& (6)& 0.8&$3.1^{+0.1}_{-0.1}$&-&   13.11&1627\\
     LE850.6&   1.541& (6)& 0.7&$3.4^{+0.3}_{-0.2}$&-&   13.08&1522\\
     LE850.8&   0.086& (6)& 0.2&$3.1^{+0.1}_{-0.5}$&-&   12.95&1141\\
    LE850.14&   1.994& (6)& 1.3&$3.9^{+0.3}_{-0.3}$&-&   13.16&1822\\
    LE850.21&   5.278& (6)& 0.8&$3.1^{+0.1}_{-0.1}$&-&   12.72& 662\\
     LE850.7&  78.395& (6)& 0.8&$3.0^{+0.1}_{-0.1}$&-&   13.30&2545\\

\end{tabular}

\caption{\label{tab:sample} Fitted parameters for the high redshift
cirrus galaxies with fixed values given in brackets. The age of the
starburst $t_*$ is fixed at 0.25 Gyr and the star formation history is
assumed to be exponentially decaying with time constant $\tau$ = 6
Gyrs. Unless otherwise stated we assume $f=1$ and
$t_m=0$. Luminosities are calculated by integrating over the whole
wavelength range covered by the models. We assume a flat Universe with
$H_o = 65$ Km/s/Mpc and $\Lambda=0.7$. Where appropriate the
luminosities (and the resulting SFR) have been corrected by the
lensing magnification factor given in the table.  }\end{table*}

Dunne \& Eales (2001) have given the results of a SCUBA mapping
program of 19 galaxies.  The far infrared spectrum of four of these
(N1614, I Zw 107, IR 1525+36, Arp 220) are obviously dominated by
a starburst component.  For a further 7 galaxies there is evidence
from the 60/850 $\mu$m colours that a starburst component is present
($log_{10} {{S(60)}\over{S(850)}} > 1.8$). We also excluded NGC7541 because
of the large flux correction applied to the 450$\mu m$ data by Dunne
\& Eales (2001).  We have therefore modelled the SEDs of the remaining
7 galaxies, including optical and near infrared data from the
literature, in terms of a cirrus component (Figure 1).  The parameters
for the models are given in Table 1. For NGC1667 we only give the
fitted parameters in Table 1. The fits are extremely good and do not
show evidence that a more elaborate cirrus model is required.  The
values of $\psi$ range from 2 to 8, and the values of $A_V$ range from
0.4 to 0.9.  Such values are consistent with the local distribution
function for the ratio $log_{10}(L_{fir}/L_{opt})$, which peaks at
-0.7 and can be fitted with a Gaussian with $\sigma$ = 0.24 about this
(Rowan-Robinson et al 1987).  As pointed out by Rowan-Robinson (1992),
the parameter $\psi$ is proportional to the mean surface brightness of
the galaxy.

\section{Application to high redshift SCUBA galaxies}

The detection of a number of galaxies in blank field surveys with
SCUBA et 850 $\mu$m (Hughes et al 1998, Eales et al 1999, Barger et al
1999, etc) and the realization that most of these are at redshift $>$
1, with some probably at z $>$ 3, has had a big impact on ideas about
galaxy formation and evolution.  Most attempts to fit the SEDs of
these galaxies (Hughes et al 1998, Downes et al 1999, Lutz et al 2001,
etc) and most attempts to fit the 850 $\mu$m counts and background
(Guiderdoni et al 1998, Dole et al 2001, Xu et al 2001, Franceschini
et al 2001, Pearson 2001, Granato et al 2001) have assumed that we
are looking at a very strongly evolving, deeply dust-embedded
starburst population.  However Rowan-Robinson (2001) argues that the
natural interpretation of the submillimetre counts and background is
in terms of a cirrus-type component.  How can the latter picture be
reconciled with the failure to identify many of the SCUBA galaxies
with optical counterparts ?  Obviously the dust optical depth must be
higher than in local disk galaxies.  However that is not an
unreasonable expectation at z = 1-3.  Calzetti and Heckman (1999) and
Pei et al (1999) have predicted how the typical dust opacity in
galaxies should evolve with epoch for star-formation histories typical
of those derived from uv and infrared surveys.  An increase of the
average $A_V$ in the interstellar medium in galaxies by a factor of 2
or 3 over present-day values at z =1-2 is to be expected, because the
increased gas-density in galaxies far outweighs the slight decline in
metallicity over the same look-back time.

Here we have selected two samples of SCUBA high-z galaxies from SCUBA
blank-field surveys, confining attention to those with the most
reliable optical identification: (a) submm galaxies which have been
confirmed by submillimetre interferometry (e.g. Lutz et al 2001), (b)
submm galaxies from the 8 mJy survey (Scott et al 2001, Fox et al
2001) which have been confirmed by association with radio sources
(Ivison et al 2002).  For both these samples the optical and near
infrared associations should be reasonably secure.  For sample (a) we
have the added bonus that the reality of the submillimetre source is
confirmed.

To make use of the radio data we have extended our model to the radio
regime by utilizing the well known far-infrared-radio correlation
(Helou et al 1985). For $\lambda > 60\mu m$ we add to the rest frame 
model SED $S_{\nu}^r$ where

$$ S_{\nu}^r = 2.7 \times 10^{-3} (2.58 S_{60\mu m} + S_{100\mu m}) \times 
    ({{\nu} \over {1.4GHz}})^{-0.8}  $$
where $S_{60\mu m}$ and $S_{100\mu m}$ are the 60 and 100$\mu m$ fluxes predicted
by the model and $\nu $ is the frequency in Hz. The factor $2.7 \times 10^{-3}$
is the maximum value allowed by the far-infrared-radio correlation (Blain 1999).

The model allows us to self-consistently estimate the star formation rate (SFR) of the
objects studied. Assuming $t_*=0.25$Gyrs the SFR is given by 

$$ SFR = {L_{bol} \over {7.88 \times 10^9}} M_\odot yr^{-1} $$

where $L_{bol}$ is the bolometric luminosity of the object (in units of solar luminosity) 
which we obtain by integrating over the whole wavelength range covered by the models.   

To test the predictions of our cirrus model we generated a grid of
models in which we vary $\psi$ (in the range 1-21 in steps of 1) and
$A_V$ (in the range 0 to 3 in steps of 0.1). For most of the sources
in these samples the redshift is unknown so we treat it as a free
parameter varying between 0 and 10 in steps of 0.1. For objects with
data in only four bands (all of the objects in Lockman East and
SMM00266+170) we fix $\psi$ to 6. For all the models we assume
$t_*=0.25 Gyrs$, $\tau=6$ Gyrs, $f=1$ and $t_m=0$.
 
Minimum $\chi^2$ fits of our cirrus model to the SEDs of these two
samples of high z SCUBA galaxies are shown in Figures 2-4, with the
parameters of the models given in Table 2. In N2 we model all the
objects with robust radio associations except N2850.5 which is a blank
field in all optical and near-IR bands, and N2850.13 which is detected
only in K.  In LE we exclude LE850.12 and LE850.18 which are detected
only in the I band. LE850.1 is included in the sample with millimetre
interferometry.  Fits that exceed the upper limits are rejected. The
quoted errors in $z$ indicate the spread in $z$ for fits with 
$\chi^2$ less than $\chi^2_{min} + 1$.

 The fits we obtain are generally good. One notable exception is
SMM02399-0136 where the high reduced $\chi^2$ is almost entirely due
to the radio point. Ivison et al. (1998) find evidence for an AGN in
this object. Also, the high fitted value of $\psi $ in this object
(the maximum value in our grid of models) is perhaps indicative of the
presence of a circumnuclear starburst accompanying the AGN in this
object. The fits to LE850.3 and LE850.7 (not plotted) are also
poor. The steep radio spectrum of these objects may also suggest the
presence of an AGN (Ivison et al 2002).

  The values of $\psi$ are higher, typically, by a factor of 2-3 than
the values for local galaxies, and the values of $A_V$ are also higher
by a factor of 2-3.  Although the values of $A_V$ are higher than in
local galaxies, they are much lower (by one or two orders of
magnitude) than would be expected in a typical molecular cloud
undergoing a starburst.  So these models are radically different from
the typical dusty starburst model for these high-redshift
submillimetre galaxies. Of course, because the bolometric
luminosities correspond to high rates of star-formation, much of the
optical and uv light which is illuminating the interstellar dust in
our cirrus models is light that has escaped from a starburst region,
for example by non-spherically symmetric evolution of the associated
HII regions. It is also interesting to note that the $A_V$ we
infer are comparable to those of molecular clouds in late stages of
their evolution (about a few times $10^7$ years) as predicted by the
starburst evolution model of ERS2000. As pointed out by the latter
authors the predicted IRAS colours of these clouds are similar to
those of cirrus galaxies. 

With a few exceptions we are able to put fairly strong constraints to
the redshifts of the sources we model. For SMMS14011+0252 Ivison et
al. (2000) report a spectroscopic redshift of 2.56 which is consistent
with our model. Smail et al. (2003) give a spectroscopic redshift
of 2.38 for N2850.4 which is  close to the deduced photometric
redshift.  For HDF850.1 there are two distinct minima, one at
$z=4.5$ and one at around $z=7.1$. In Table 2 we give the fitted
parameters for both models.  For LE850.2 there is a broad minimum
between $z=4$ and $9$.

Of the 23 objects in our samples, our models successfully fit the SEDs
of 16 of them (70 $\%$). For 4 objects (17$\%$) we have insufficient
data (fewer than 4 bands) to be able to test our model. For these
objects we may need a higher optical depth model of the M82 or Arp 220
starburst type or values of $A_V$ ($>3$) from our cirrus model which
are not really consistent with the '$A_V$ not $>>$ 1' requirement of
our model. For 3 objects our fitted $\psi $ is close to the maximum in
the range considered (21).  We assumed the age of the starburst $t_*$
= 0.25 Gyr in each case.  This will underestimate the contribution of
older, low-mass stars in the rest-frame near infrared if present.  In
the case of N2850.8 we do see evidence for such a near-IR excess.

\section{Discussion and conclusions}

Although the current state of observations of high-z SCUBA galaxies is
insufficient to tell whether our models are better or worse than
starburst models, the fits shown here demonstrate that it is premature
to make an analogy between SCUBA galaxies and dusty ultraluminous
starburst galaxies like Arp 220.  Quite a modest dust optical depth
($A_V \sim 1-3$) combined with the redshifting of the optical and uv
radiation is sufficient to explain why these galaxies are faint in the
optical.

The inferred bolometric luminosities imply that enhanced star
formation has taken place in these galaxies, so even if we are correct
that the observed submillimetre emission arises mainly from cirrus
emission, ie reemission from dust in the general interstellar medium
of the galaxy, it is likely that some optical and ultraviolet light is
intercepted by dust in the dense molecular gas from which the stars
formed.  Such emission would peak at $\sim$ 50 (1+z) $\mu$m and could
easily contribute bolometric luminosities comparable to the cirrus
luminosities of table 2 without being detectable at present.
Estimates of the fraction of optical-uv light which escapes from
star-forming regions in the local universe range from 25-50 $\%$.  In
Fig 4 we have also shown fits for a more elaborate model in which
newly formed stars are completely shrouded in their molecular cloud
for the first $10^7$ yrs, 50$\%$ of the starlight escapes to
illuminate the interstellar dust from $10^7 - 7.2 \times 10^8$ yrs
(i.e. $f=0.5$ and $t_m=10^7$ years) and 100$\%$ escapes thereafter (cf
the starburst evolution models of ERS2000). The starlight that escapes
is reprocessed to the infrared in the same way as the pure cirrus
model. We have generated a grid of models spanning the same parameter
space in $A_V$, $\psi$ and $z$ as for the pure cirrus model. These
models, when observed at 0.25 Gyr from the start of the starburst, are
very similar to our pure cirrus models except for additional radiation
peaking at 50 $\mu$m.

A critical test of our models would be to measure the angular extent
of the millimetre or submillimetre sources in high-z galaxies.  In our
model these should be of order an arcsec, within range of current
millimetre-wave interferometers, whereas the obscured nuclear
starburst model would imply sizes an order of a magnitude smaller. Of
course, this test would not be able to discriminate between our model
and a model where star formation takes place in an extended highly
obscured starburst.

There is a further factor at work which exaggerates the difference
between the SCUBA galaxies and the high-redshift galaxies found in
Lyman break surveys (Steidel et al 1999).  Assuming that there is a
spread in the typical optical depth of the interstellar medium, uv
surveys will inevitably be biased towards galaxies with lower than
average optical depth, while submm galaxies will be biased towards
higher optical depth.  According to this picture, if the Lyman break
surveys could go a factor 10 deeper they would start to see the SCUBA
galaxies, and if the SCUBA surveys could go a factor 10 deeper they
would start to see the Lyman break galaxies.  This is already hinted at
by the analysis of Peacock et al (2000).

\section*{Acknowledgements} 

This research has made use of the NASA/IPAC Extragalactic Database
(NED) which is operated by the Jet Propulsion Laboratory, California
Institute of Technology, under contract with the National Aeronautics
and Space Administration. The authors are grateful to Steve Warren,
Jose Afonso, Stephen Serjeant, Gian Luigi Granato, Toshi Takagi,
Dieter Lutz and Alberto Franceschini for their useful comments on an
earlier version of this paper. We would also like to thank the
anonymous referee for constructive comments and suggestions. Last, but
not least, we would like to thank our collaborator Ralf Siebenmorgen
for making his code for computing the emission from small grains and
PAHs available to us. AE acknowledges support by PPARC.

\end{document}